\documentclass[prl,twocolumn,showpacs,preprintnumbers,amsmath,amssymb,nofootinbib]{revtex4}

\usepackage{amsmath}
\usepackage{amsfonts}
\usepackage{amssymb}

\newcommand{\be}{\begin{equation}}
\newcommand{\ee}{\end{equation}}
\newcommand{\bear}{\begin{eqnarray}}
\newcommand{\eear}{\end{eqnarray}}
\newcommand{\beqstar}{\begin{eqnarray*}}
\newcommand{\eeqstar}{\end{eqnarray*}}

\begin{document}

\preprint{HUTP-04/A032}

\title{Single field consistency relation for the 3-point function}

\author{Paolo Creminelli$^*$ and Matias Zaldarriaga$^{* \dagger}$}

\affiliation{$^*$Jefferson Physical Laboratory, Harvard University, Cambridge, MA, 02138, USA \\
$^\dagger$Astronomy Department, Harvard University, Cambridge, MA, 02138, USA }

\begin{abstract}
We point out the existence of a consistency relation involving the 3-point function of scalar perturbations which is valid in any inflationary
model, independently of the inflaton Lagrangian under the assumption that the inflaton is the only dynamical field. The 3-point function in the limit in which 
one of the momenta is much smaller than the other two is fixed in terms of the power spectrum and its tilt. This relation, although very hard to verify
experimentally, could be easily proved wrong by forecoming data, thus ruling out any scenario with a single dynamical field in a model independent way.

\end{abstract}


\maketitle

In the last few years various modifications of the single field slow-roll inflation scenario have been proposed. The basic mechanism
for solving the standard cosmological problems remains unaltered: a rapid expansion with an approximately constant Hubble parameter.
What is different in these recent models are the characteristics of the produced density perturbations. These are modified if we change
the inflaton dynamics with respect to the minimal slow-roll case, for example with higher derivative terms
in the inflaton Lagrangian \cite{Armendariz-Picon:1999rj,Creminelli:2003iq,Arkani-Hamed:2003uz,Alishahiha:2004eh} or adding sharp 
features in the inflaton potential \cite{Adams:2001vc}. Another possibility is to assume that density perturbations are created by another field, 
different from the inflaton, whose quantum fluctuations are finally converted into adiabatic perturbations \cite{Lyth:2001nq,Dvali:2003em}.

These alternative models generically give distinctive predictions for the shape of the scalar spectrum, the gravitational wave contribution and
for the scalar 3-point function, which should allow to distinguish them from a minimal slow-roll model.
Without much theoretical guidance about which kind of modification is more likely, it would be extremely useful to make model independent
statements about the observable quantities.

The purpose of this letter is to indicate the existence of a consistency relation involving the 3-point function of scalar perturbations, which is valid
in any inflationary model {\em irrespectively of the inflaton dynamics} (\footnote{A different kind of consistency relation valid in a 
particular set of models has been recently pointed out in \cite{Gruzinov:2004jx}.}), under the assumption that the inflaton is the only dynamical field
during inflation. It is not based on any slow-roll approximation and it is valid for any inflaton Lagrangian. We will show that the consistency 
relation is in some sense kinematical, being just a consequence of the assumption that there are no other fields which evolve (classically or 
quantum mechanically) during inflation. The inflaton is the only ``clock of the Universe" and it fixes the Hubble parameter: fluctuations
of the inflaton are therefore equivalent to a relative rescaling of the scale factor in different parts of the Universe. 

The consistency relation relates a particular geometrical limit of the 3-point function of density perturbations to the spectrum and tilt of the 
2-point function:
\be
\label{eq:cons}
\lim_{k_1 \rightarrow 0} \langle\zeta_{\vec \kappa_1}\zeta_{\vec \kappa_2}\zeta_{\vec \kappa_3}\rangle =
- (2\pi)^3 \delta^3 (\sum_i \vec k_i) (n_s-1) P_{k_1} P_{k_3} \;.
\ee
On the left-hand side we have the 3-point function of the $\zeta$ variable, which is the non-linear generalization of the well known
variable introduced by Bardeen, Steinhardt and Turner \cite{Bardeen:qw}. The non-linear generalization, introduced in \cite{Salopek:1990jq}, 
is such that the metric, once a mode is frozen outside the horizon, can be written as
\be
\label{eq:zeta2}
ds^2 = -dt^2 + e^{2 \zeta(x)} a(t)^2 dx_i dx_i \;. 
\ee
The 3-point function is evaluated in the particular geometrical limit in which one of the wavevectors ($\vec k_1$) becomes much smaller
than the other two. For momentum conservation this implies that the other two momenta become equal and opposite. 
On the right-hand side we have the two power spectra of the $\zeta$ variable, one for the long mode $k_1$ and one for the short modes 
$k_2 \simeq k_3$. The power spectrum is defined by
\be
\label{eq:power}
\langle\zeta_{\vec k_i}\zeta_{\vec k_j}\rangle = (2\pi)^3 \delta^3 (\vec k_i +\vec k_j) P_{k_i} \;,
\ee
while $n_s-1$ is the usual tilt of the scalar spectrum: $\langle\zeta\zeta\rangle \sim k^{-3+(n_s-1)}$. If the variation of the scalar tilt is not
negligible $n_s$ must be considered a function of $k$. In this case $n_s$ must be replaced by $n_s(k_3)$ in the right-hand side of 
the consistency relation (\ref{eq:cons}).

Let us now proceed to the proof of eq.~(\ref{eq:cons}), which will underline which are the assumptions we have to make. In an 
inflating Universe, independently of the specific inflaton dynamics, modes with longer wavelength freeze earlier\footnote{A mode will freeze when its 
frequency is of order $H$. This in general does not imply that its wavelength is of the order of the Hubble 
horizon unless the inflaton fluctuations have a standard dispersion relation with a sound speed equal to the speed of light. This is true
in standard slow-roll inflation, but it is not the case in more generic models \cite{Armendariz-Picon:1999rj,Arkani-Hamed:2003uz,Alishahiha:2004eh}.}.
Therefore, taking $k_1 \ll k_{2,3}$, the first mode will be already frozen outside the horizon when the two smaller ones freeze.
As we are assuming that there is only one clock, the inflaton value fixes the Hubble parameter so that the only difference among
different parts of the Universe is a rescaling of the spatial coordinates as in eq.~(\ref{eq:zeta2}). It is important to note that we are implicitly 
assuming that the perturbations are just fluctuations back and forth on the {\em same unique classical solution}; at the same value of the 
inflaton $\phi$ we have the same value of $\dot\phi$, $H$ and so on. This is the case if the classical solution is a dynamical attractor, so that 
only one mode of the perturbations remains after horizon crossing while the other decays exponentially. Thus to make our assumptions more precise: we are assuming that, 
neglecting decaying modes, only a single classical evolution is possible, so that the only allowed perturbations are along this single solution and
they can thus be written in terms of the $\zeta$ variable as in eq.~(\ref{eq:zeta2}). In other words we do not have any isocurvature component. 

The 2-point function $\langle\zeta_{\vec k_2}\zeta_{\vec k_3}\rangle$ will now depend on the value of the background wave $\zeta_1$ already 
frozen outside the horizon. As it is clear from the metric (\ref{eq:zeta2}), a long wavelength mode $\zeta_1$ is equivalent, 
neglecting gradients, to a rescaling of the coordinates. In position space the variation of the 2-point function given by a classical 
background $\zeta_1$ is at linear order
\be
\label{eq:derivative}
\frac{\partial}{\partial \zeta_1} \langle\zeta(x)\zeta(0)\rangle \cdot \zeta_1 = x \frac{d}{d x} \langle\zeta(x)\zeta(0)\rangle \cdot \zeta_1 \;.
\ee
To get the 3-point function we have to multiply the expression above by $\zeta_1$ and average over it. Going to Fourier space leads to our 
consistency relation (\ref{eq:cons}). Again in the assumption that the inflaton is the only clock, $\zeta$ is constant 
outside the horizon \cite{Salopek:1990jq}, because neglecting gradients the effect of $\zeta$ is equivalent to an unobservable rescaling of the 
coordinates: every observer goes through exactly the same history. This implies that the consistency relation, obtained 
during inflation, holds later whatever is the evolution of the Universe, before perturbations finally reenter in the 
horizon. Note that the consistency relation holds only if we generalize the linear expression $(1+2\zeta(x))$ to the exponential form used in the metric
(\ref{eq:zeta2}), although the variable $\zeta$ would be conserved outside the horizon even using another form. The reason is that only with the exponential form
the effect of the background wave $e^{\zeta_1}$ multiplies the small scale fluctuations and can therefore be traded with a scale redefinition on the short modes.

Eq.~(\ref{eq:cons}) was derived by Maldacena \cite{Maldacena:2002vr} in his calculation of the 3-point function in single-field slow-roll inflation. 
Our point here is to stress that this consistency relation has a much broader validity, being only a consequence of the assumption that the inflaton 
is the only clock in the Universe. The effect of a long wavelength mode is just a scale redefinition so that the 3-point function is fixed in the limit
$k_1 \ll k_{2,3}$ and vanishes if the 2-point function is scale invariant. The reader might be surprised that the scalar spectrum, which can be calculated
from the quadratic Lagrangian of the fluctuations, knows something about the 3-point function, which is an intrinsically non-linear object. In the limit
we are considering the additional information is encoded in the background cosmological solution $a(t)$ as the long wavelength mode is equivalent to an 
overall rescaling of the unperturbed history $a(t)$. 

Experimental limits on non-gaussianity are set on the scalar variable $f_{\rm NL}$ defined through the relation
\begin{equation}
\label{eq:f_NL}
\zeta(x) =\zeta_g(x) -\frac35 f_{\rm NL}(\zeta_g(x)^2 - \left<\zeta_g^2\right>) \;,
\end{equation}  
where $\zeta$ is the observed perturbation and $\zeta_g$ is gaussian\footnote{The non-linearity parameter is defined for the Newtonian potential in 
matter dominance; this explains the factor of 3/5 in the definition.}. This assumes that the non-gaussianity is introduced by 
quadratic corrections at the same point in space. The present limits on $f_{\rm NL}$ are given by the WMAP experiment $-58 < f_{\rm NL} < 134$ at $95\%$ of 
CL \cite{Komatsu:2003fd}. The Planck experiment will improve this limit down to $f_{\rm NL} \sim 5$. 
Definition (\ref{eq:f_NL}) implies a 3-point function which reduces in the squeezed limit $k_1 \ll k_{2,3}$ to 
\be
\label{fNLlimit}
\lim_{k_1 \rightarrow 0} \langle\zeta_{\vec \kappa_1}\zeta_{\vec \kappa_2}\zeta_{\vec \kappa_3}\rangle =
(2\pi)^3 \delta^3 (\sum_i \vec k_i) \; 4 f_{\rm NL} P_{k_1} P_{k_3} \;.
\ee
From this expression we obtain that our consistency relation implies a {\em very low level} of non-gaussianity in the particular geometrical
limit $k_1 \ll k_{2,3}$, much below current and forecoming limits. For single field slow-roll inflation the 3-point function remains small
even considering triangles in momentum space with comparable sides \cite{Maldacena:2002vr,Acquaviva:2002ud}. This, however, is not true in general for any 
model where the inflaton is the only dynamical field. Various models have been proposed which predict quite a big level of non-gaussianity, not far 
from the present limits \cite{Creminelli:2003iq,Arkani-Hamed:2003uz,Alishahiha:2004eh}. For these models the 3-point function, which is 
big for generic triangles, becomes very small in the limit $k_1 \ll k_{2,3}$.

The reader might be worried about the interest of the proposed consistency relation. It implies that the 3-point function is very small in a given 
geometrical limit, so that the relation will probably never be verified. The point is that it could be easily {\em proved wrong}, therefore ruling out 
in a model independent way the possibility that the inflaton is the only dynamical field. Models where density perturbations are created by 
fluctuations in a second field (as a curvaton \cite{Lyth:2001nq} or a field which changes the inflaton decay width \cite{Dvali:2003em}) obviously do 
not respect the consistency relation; fluctuations in $\zeta$ are sourced outside the horizon and the whole picture is different 
\footnote{The violation of the consistency relation does not require the existence of a second field which fluctuates during inflation. 
Even a field which evolves classically with negligible quantum fluctuation (for example if its mass is much bigger than H) introduces a second 
independent clock, so that fluctuations in the inflaton field are not equivalent to a coordinate rescaling. Obviously this second field 
must be really independent and not triggered by the inflaton itself as it happens, for example, in an hybrid inflation model.}. As non-linearities 
develop outside the horizon these scenarios generically predict non-gaussianities which are local in position space as in eq.~(\ref{eq:f_NL}). 
The  value of $f_{\rm NL}$ is model dependent but it has a rough lower bound at the level of what Planck can detect \cite{Zaldarriaga:2003my,Lyth:2002my}. 
The detection of a 3-point function of this kind would {\em experimentally rule out} the proposed consistency relation and with it any model 
with a single dynamical field, independently of its dynamics. The result would have the same implications as the detection of an isocurvature component 
in the scalar perturbations.

In order to rule out the proposed consistency relation, data should be able to characterize the dependence of the 3-point function on the shape
of the triangle in momentum space to isolate the limit $k_1 \ll k_{2,3}$. The relation tells us that the 3-point function is very small in the limit
$k_1 \rightarrow 0$, but in order to make it a testable statement we must say something about the subleading corrections in this limit. Corrections
come from the dependence of the 2-point function on {\em derivatives} of the background wave $\zeta_{\vec k_1}$. Terms proportional to the first
derivative of the long wavelength mode do not contribute to the 3-point function because their effect averages to zero 
$\langle\zeta_{\vec k_1} \nabla \zeta_{\vec k_1}\rangle =0$. Subleading corrections in the limit $k_1 \rightarrow 0$ are therefore suppressed by
$(k_1/k_3)^2$ giving for $k_1 \ll k_{2,3}$
\be
\label{eq:subleading}
\langle\zeta_{\vec \kappa_1}\zeta_{\vec \kappa_2}\zeta_{\vec \kappa_3}\rangle \simeq
- (2\pi)^3 \delta^3 (\sum_i \vec k_i) (n_s-1+{\cal{O}}\left(\frac{k_1}{k_3}\right)^2) P_{k_1} P_{k_3} \;.
\ee   
Thus the relevant experimental question is the following. Imagine that the level of non-gaussianity is sufficiently big to be observable. Can we distinguish models which satisfy
the consistency relation (\ref{eq:subleading}), whose 3-point function dies at least as fast as $(k_1/k_3)^2$ in the squeezed limit, from models which violate the consistency
relation, for example with a non-gaussianity of the local form (\ref{eq:f_NL})? The comparison between these two classes of models have been 
studied in \cite{Babich:2004gb}. The conclusion is that these different dependences on momenta of the 3-point function give experimental signals
that are quite ``orthogonal" when a cosine between distributions based on signal to noise weighting is used\footnote{Only the cosmic
variance noise is used to define this cosine. Obviously if detector noise is large it can prevent the detection of the non-Gaussian signatures. 
Distinguishing between different configuration dependences is obviously more difficult than detecting a signal of non-gaussianity.}. To be more explicit,
if our Universe has for example a non-Gaussianity of the local form (\ref{eq:f_NL}), a global fit to the data assuming the local model will decrease 
the $\chi^2$ with respect to the gaussian case by $\Delta\chi^2_{\rm local}$. If the data are fitted assuming a different distribution 
the $\chi^2$ will only decrease by a factor $\Delta\chi^2_{\rm local} \cos^2\theta$. Typical values of $\cos\theta$ obtained in \cite{Babich:2004gb}
between the local distribution and one satisfying the consistency relation are of order $0.5 \div 0.6$ for a CMBR experiment. 

Another important point must be stressed regarding the experimental possibility to observe a deviation from the proposed consistency relation.
In real experiments we do not observe directly $\zeta$ but some physical quantity, as the temperature fluctuations of the CMBR, which
in first approximation is linearly related to $\zeta$. Non-linear corrections introduce non-gaussianity in the quantity we observe,
even starting from a completely gaussian perturbation $\zeta$. These effects are a sort of background if we are interested in the 3-point
function of $\zeta$. Generic second order effects will introduce corrections with $f_{\rm NL} \sim 1$, which should be below the sensitivity 
of the Planck experiment, so that the measurement should not be ``contaminated'' by non-linearities. Explicit calculations in specific limits 
\cite{Bartolo:2003bz,Creminelli:2004pv} confirm this expectation, although the full non-linear calculation relating $\zeta$ to the final temperature 
perturbation we observe has not been carried out. In particular the calculation for the squeezed limit $k_1 \ll k_{2,3}$ was done including all relevant 
effects only for the case when the long wavelength mode was outside the horizon at recombination \cite{Creminelli:2004pv}.

In the discussion above we have not introduced gravity waves. Do they represent a second clock, thus changing the whole picture? As we discussed, in the presence
of a single dynamical field, different regions outside the horizon behave as ``parallel Universes'': every observer goes exactly through the same history and 
the only difference is an unobservable spatial coordinate rescaling. But tensor modes represent a second fluctuating mode, can they act as a curvaton and 
change $\zeta$ outside the horizon?  The answer in no, they cannot. It is easy to prove that both $\zeta$ and the tensor component are constant at any 
perturbative order outside the horizon \cite{Salopek:1990jq,Maldacena:2002vr}. In fact, both $\zeta$ and the tensor contribution $h$ (with $\det h =1$) in the metric 
\be
\label{eq:GW}
ds^2 = -dt^2 + e^{2 \zeta(x)} h_{ij} a(t)^2 dx^i dx^j  
\ee
can be reabsorbed into an unobservable rescaling of coordinates outside the horizon, so that they are separately conserved. In practice the ``parallel Universes''
point of view can be generalized even in the presence of a gravity wave contribution: every observer still goes through the same history and scalar and tensor modes 
have physical meaning only when they reenter in the horizon.

The consistency relation (\ref{eq:cons}) does not apply to ``bouncing'' models as the ekpyrotic-cyclic scenarios \cite{Khoury:2004xi}, although
they are based on a single scalar field. To prove the relation we have assumed that a quantum fluctuation of the long mode can be interpreted, after
horizon crossing, as a classical $\zeta$ wave.
As pointed out for example in \cite{Steinhardt:2004gk}, $\zeta$ is a decaying mode in the contracting phase of the ekpyrotic-cyclic scenarios, so that it
is not possible to interpret a mode after exiting the horizon as a classical $\zeta$ wave. For the same reason it is not possible to prove that 
$\zeta$ is constant at the bounce. 

In conclusion, we have pointed out a consistency relation which is valid under the only assumption that a shift of the inflaton is equivalent to an 
overall rescaling of the spatial coordinates.
If the level of non-gaussianity is sufficiently high, new data could experimentally rule out the proposed relation. This would be a model independent
proof of the presence of more than one dynamical field in the early Universe, quite similarly to the detection of an isocurvature component in the scalar
perturbations.
 
{\em Acknowledgments}. We would like to thank Leonardo Senatore, Toby Wiseman and especially Alberto Nicolis for useful discussions. 
M.~Z.~ is supported by NSF grants AST 0098606 and by the David and Lucille Packard Foundation Fellowship for Science and Engineering.


\end{document}